\documentclass[a4paper]{article}
\usepackage{INTERSPEECH2019}
\usepackage{multirow}
\usepackage{array}
\usepackage{multicol}
\usepackage{hyperref}
\title{Unsupervised Speech Domain Adaptation Based on Disentangled Representation Learning for Robust Speech Recognition}
\name{Jong-Hyeon Park$^1$, Myungwoo Oh$^2$,  Hyung-Min Park$^1$}
\address{
  $^1$Department of Electronic Engineering, Sogang University, Seoul 04107, Republic of Korea\\
  $^2$Clova AI Research, Naver Corp., Gyeonggi-do 13561, Republic of Korea}
\email{vidovic@sogang.ac.kr, myungwoo.oh@navercorp.com, hpark@sogang.ac.kr}
\begin{document}
\maketitle
\begin{abstract}
In general, the performance of automatic speech recognition (ASR) systems is significantly degraded due to the mismatch between training and test environments. Recently, a deep-learning-based image-to-image translation technique to translate an image from a source domain to a desired domain was presented, and cycle-consistent adversarial network (CycleGAN) was applied to learn a mapping for speech-to-speech conversion from a speaker to a target speaker. However, this method might not be adequate to remove corrupting noise components for robust ASR because it was designed to convert speech itself.
In this paper, we propose a domain adaptation method based on generative adversarial nets (GANs) with disentangled representation learning to achieve robustness in ASR systems. In particular, two separated encoders, context and domain encoders, are introduced to learn distinct latent variables. The latent variables allow us to convert the domain of speech according to its context and domain representation. We improved word accuracies by 6.55\texttildelow15.70\% for the CHiME4 challenge corpus by applying a noisy-to-clean environment adaptation for robust ASR. In addition, similar to the method based on the CycleGAN, this method can be used for gender adaptation in gender-mismatched recognition.

\end{abstract}

\section{Introduction}
Automatic speech recognition (ASR) is getting more and more attention because speech can provide the most user-friendly interface for smart devices. Unfortunately, an input speech signal is corrupted by noise in most of practical situations. With the corrupted speech, the performance of ASR systems is significantly degraded due to the mismatch between training and test environments. Therefore, noise robustness remains a very important issue in the field of ASR.
Especially, in case of the ASR systems with acoustic models based on deep learning, training data acquired in various environments help to alleviate the mismatch, but it is very difficult to obtain sufficient amounts of such data. Many methods have been proposed to achieve robustness of ASR systems even in the case such data cannot be sufficiently obtained~(e.g.,~\cite{anguera2007acoustic,krueger2010model,seltzer2004bayesian,cho2017bayesian}).

Since deep learning emerged as a breakthrough for acoustic modeling, it has also been applied to speech enhancement or preprocessing for robust ASR~(e.g.,~\cite{heymann2016neural,lu2013speech,narayanan2013ideal,lee2016dnn}). Recently, domain\footnote{The domain in speech may mean a speaker uttering the speech or a situation or an environment in which the speech is uttered.} adaptation methods that convert input domains to training domains were presented to alleviate the mismatch between the training and test domains. In particular, several deep-learning-based domain adaptation methods, such as the methods based on variational auto-encoder~\cite{DBLP:journals/corr/HsuZG17aa} and adversarial leaning~\cite{denisov2018unsupervised}, were proposed for robust ASR in noisy environments by learning the mapping between noisy and clean speech. In addition, a deep-learning-based image-to-image translation technique to translate an image from a source domain to a desired domain was presented, and cycle-consistent adversarial network (CycleGAN)~\cite{zhu2017unpaired} was applied to learn a mapping for speech-to-speech adaptation from a speaker to a target speaker for gender-mismatched recognition~\cite{hosseini2018multi}. The CycleGAN-based domain adaptation method attracted interest because it might be easier to train its model than the others mentioned before. However, this method might not be adequate to remove corrupting noise components for robust ASR because it was designed to convert speech itself.

In this paper, we propose a domain adaptation method based on generative adversarial nets (GANs)~\cite{goodfellow2014generative} with disentangled representation learning to achieve robustness in ASR systems. With this method, the model can separately learn domain mappings between different noise conditions as well as speakers. As an image style translation model based on disentangled representation learning, the MUNIT model divided the encoder in the generator into two, and each encoder generated latent variables presenting different attributes of an input image such as content or style~\cite{huang2018multimodal}. In addition, it exploited the adaptive instance normalization method~\cite{huang2017arbitrary} to apply latent variables for generation of improved style-converted images. Similar to the MUNITs, we compose two separated encoders, context and domain encoders, in our proposed model. Both encoders generate latent variables that represent different attributes, by forcing the latent variables to have different prior probabilities. The learned model can generate mel-spectrograms of speech adapted to a desired domain by combining the latent variables for the two separated encoders with the adaptive instance normalization method. The model is composed of convolutional neural networks (CNNs), residual networks~\cite{he2016deep}, and fully-connected neural networks (FCNNs). In Section 2, we describe the conventional CycleGAN method and explain how to adapt the speech domain with this method. Then, we describe our method that includes the disentangled representation learning with the adaptive instance normalizations method. The model is evaluated by ASR experiments in Section 3, and concluding remarks are summarized in Section 4.

\section{Proposed method}
\subsection{Review of the CycleGAN}
CycleGAN is an image-to-image translation method based on GANs, which can perform conversion between unpaired images~\cite{zhu2017unpaired}. The GANs is a generative model composed of two networks, a generator, and a discriminator. The generator aims to generate an image similar to training images by mapping a distribution of training images from a random distribution. The discriminator aims to discriminate that the generated image is true or not. Several studies have suggested effective image translation methods based on the GANs. Common networks effective for this image-style translation are DualGAN~\cite{yi2017dualgan}, DiscoGAN~\cite{kim2017learning} and CycleGAN. All of these models have similar networks, so we will focus on the CycleGAN network and we apply it for speech domain adaptation.

The CycleGAN consist of two pairs of generators ($G_{XY}$, $G_{YX}$) and discriminators ($D_{X}$, $D_{Y}$). Each generator and discriminator is applied to either input or target domain. The goal of the model is to allow a generator to act as a converting function to either domain. To make it work, the CycleGAN defines two types of loss functions. One is the adversarial loss function that is the same as that of the GANs. The adversarial loss function is defined as follows:
\begin{equation}
\label{eq1}
\begin{aligned}
  \mathcal{L}_{adv}(G_{XY}, D_{Y})&= \mathbb{E}_{ y \sim P_{data(y)}  } [ \log D_{Y}(y)] \\
  					& +\mathbb{E}_{ y \sim P_{data(x)}  } [ \log ( 1-D_{Y}( G_{XY}(x))].
\end{aligned}
\end{equation}
The other is called the cycle-consistency loss. The cycle-consistency loss is defined by getting the mean-square error between the input and its output reconstructed by applying two generators, $G_{XY}$ and $G_{YX}$. The loss reduces the space of possible mapping functions and ensures that the learning remains cycle-consistent. The cycle-consistency loss function is defined as
\begin{equation}
\begin{aligned}
  \mathcal{L}_{cyc}(G_{XY}, G_{YX})&= \mathbb{E}_{ x \sim P_{data(x)}  } [ \Vert G_{YX}( G_{XY}(x)) - x \Vert _{1} ] \\
  					& +\mathbb{E}_{ y \sim P_{data(y)}  } [ \Vert G_{XY}( G_{YX}(y)) - y \Vert _{1} ],
    \label{eq2}
\end{aligned}
\end{equation}
where $\Vert\cdot\Vert_1$ denotes the L1 norm.
With a cycle-consistency hyperparameter $\lambda _{cyc}$, the overall loss function is
\begin{equation}
\begin{aligned}
  \mathcal{L}&= \mathcal{L}_{adv}(G_{XY}, D_{Y})+\mathcal{L}_{adv}(G_{YX}, D_{X}) \\
  &+\lambda _{cyc}\mathcal{L}_{cyc}.
    \label{eq3}
\end{aligned}
\end{equation}

For more effective application on speech, we may apply the skip-connections used in the U-net~\cite{ronneberger2015u} to ensure that linguistic information is maintained as much as possible. We use mel-spectrogram segments as input features, and more detail will be described in Section 3.

\subsection{Disentangled representation learning with adaptive instance normalization}
Here, we present a domain adaptation method based on disentangled representation learning. When the CycleGAN-based method is used for noisy-to-clean speech domain adaptation, speech information may be distorted during the training because the model learns a domain mapping based on the ``pixel-wise reconstruction'' loss function. Therefore, it makes hard to get proper results on domain adaptation of situational or environmental mismatches such as noisy-to-clean speech domain adaptation. To overcome this problem, we introduce disentangled representation learning to the speech domain adaptation. Like the MUNIT model that is used for image style translation based on the disentangled representation learning,
the proposed model has two divided encoders in the generator. One encoder generates attributes that should not be changed during conversion, and the other encoder generates attributes that should be changed, such as environmental conditions or speaker characteristics. In this paper, we call these attributes as context and domain, respectively. The operations of the context and domain encoders are
\begin{equation}
\begin{aligned}
 &Enc_{A}^{c}(x_{A})=z_{A}^{c} ,~~~  Enc_{A}^{d}(x_{A})=z_{A}^{d}, \\
 &Enc_{B}^{c}(x_{B})=z_{B}^{c} ,~~~  Enc_{B}^{d}(x_{B})=z_{B}^{d},
  \label{eq4}
  \end{aligned}
\end{equation}
where $x_{A}$, $z_{A}$, and $Enc_A$ denote an input feature of speech $A$ (frequently given by a segment of mel-spectrograms), its latent variable vector, and its encoder, respectively. Superscripts $c$ and $d$ mean context and domain attributes.
\begin{figure}
  \centering
  \includegraphics[width=\columnwidth]{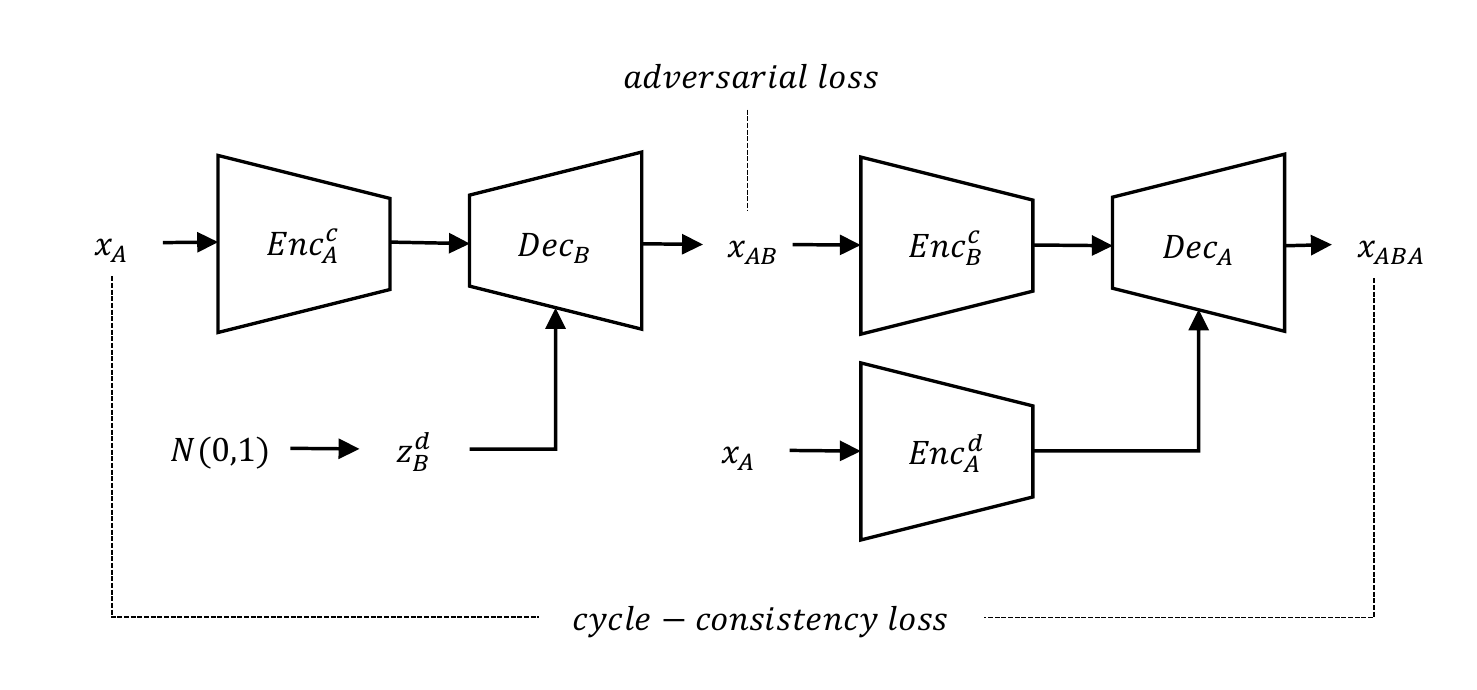}
  \caption{Proposed model with adversarial and cycle-consistnecy losses. It shows a diagram when $ x_ {A} $ is an input, but a similar diagram may be applied for $ x_ {B} $.}
 \label{figure1}
     \vspace{-0.43cm}
\end{figure}
\begin{figure}
  \centering
  \includegraphics[width=\columnwidth]{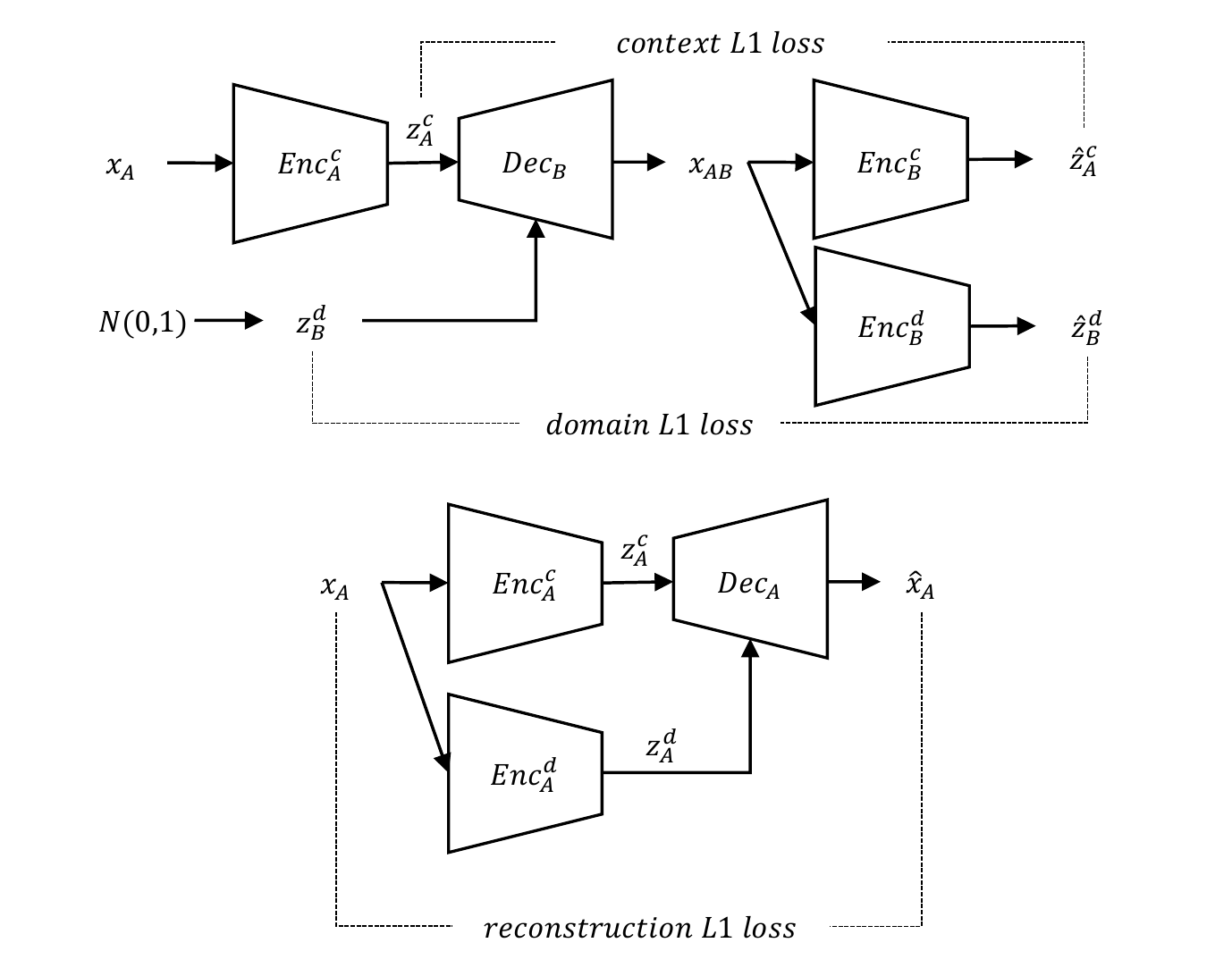}
  \caption{Proposed model with reconstruction losses. It shows a diagram when $ x_ {A} $ is an input, but a similar diagram may be applied for $ x_ {B} $.}
  \label{fig:speech_production}
    \label{figure2}
    \vspace{-0.43cm}
\end{figure}

Then, the decoders in the generator, $Dec_{A}$ and $Dec_{B}$, aim to generate converted speech features that maintain the context attributes of input speech in a domain of target speech by using context latent variables ($z_{A}^{c}$, $z_{B}^{c}$) from the input speech and domain latent variables ($z_{A}^{d}$, $z_{B}^{d}$) from the target speech. The outputs of the decoders are determined by
\begin{equation}
\begin{aligned}
 &Dec_{A}( z_{B}^{c} , z_{A}^{d}  ) = x_{BA}, &Dec_{B}( z_{A}^{c} , z_{B}^{d}  ) = x_{AB}, \\
 &Dec_{A}( z_{A}^{c} , z_{A}^{d}  ) = \hat x_{A},&Dec_{B}( z_{B}^{c} , z_{B}^{d}  ) = \hat x_{B}.
\label{eq5}
\end{aligned}
\end{equation}
$x_{AB}$ denotes a converted feature that preserves context attributes in $x_A$ with domain attributes in $x_B$, and $\hat x_{A}$ means a reconstructed feature that preserves both context and domain attributes in $x_{A}$. Discriminators $D_{A}$ or $D_{B}$ aim to discriminate real $x_A$ and its fake $x_{BA}$ or real $x_B$ and its fake $x_{AB}$. Figures~\ref{figure1} and~\ref{figure2} display the proposed model and losses to train it, respectively. Similar to the CycleGAN, we define adversarial and cycle-consistency loss functions. The loss functions for real $x_A$ and its fake one are as follows:
\begin{equation}
\begin{aligned}
\mathcal{L}_{A~adv}& = \mathbb{E} [ \log D_{A}(x_{A})], \\
&+\mathbb{E} [ \log ( 1-D_{A}( Dec_{A}(   Enc_{B}^{c}(x_{B}),  z_{A}^{d} )  ) )].
    \label{eq6}
\end{aligned}
\end{equation}
\begin{equation}
\begin{aligned}
&\mathcal{L}_{A~cyc}=\mathbb{E} [ \Vert Dec_{A}( Enc_{B}^{c}(x_{AB}), Enc_{A}^{d}(x_{A})) - x_{A} \Vert _{1} ]. \\
   \label{eq7}
\end{aligned}
\end{equation}
Also, we define additional loss functions regarding reconstructions to train paired encoders and decoders. As shown in Fig.~\ref{figure2}, the reconstruction losses are defined as the L1 norms of speech features and latent variables, which allow encoders and decoders to learn consistent mapping functions. The loss functions for $x_A$ is defined as
\begin{equation}
\begin{aligned}
 &\mathcal{L}_{A~recon}^{c}= \mathbb{E} [ \Vert   Enc_{B}^{c}(Dec_{B}( z_{A}^{c}, z_{B}^{d}  )) -  z_{A}^{c}  \Vert _{1} ], \\
 &\mathcal{L}_{A~recon}^{d}= \mathbb{E} [ \Vert   Enc_{A}^{d}(Dec_{A}( z_{B}^{c}, z_{A}^{d}  )) -  z_{A}^{d}  \Vert _{1} ], \\
 &\mathcal{L}_{A~recon}^{feat}= \mathbb{E} [ \Vert   Dec_{A}( z_{A}^{c} , z_{A}^{d}  ) -  x_{A}  \Vert _{1} ]. \\
    \label{eq8}
\end{aligned}
\end{equation}

For more effective disentangled representation learning, we apply adaptive instance normalization originally used in image-style transfer. In this approach, instance normalization has no learnable affine parameters. Instead, it receives its affine parameters from domain latent variables. Thus, we can expect domain adaptation in speech features by using domain latent variables representing statistics of speech features. Also, we assume the prior of the domain latent variables as the zero-mean unit-variance Gaussian distribution. The domain latent variables in the learning process are obtained by sampling from the prior. For simple derivation, the domain latent variables are learned to follow the zero-mean unit-variance Gaussian distribution by using the second equation in Eq.~(\ref{eq8}), instead of the Kullback-Leibler divergence. Finally, the total loss function is defined as follows:
\begin{equation}
\begin{aligned}
 \mathcal{L}_{total}&= \mathcal{L}_{adv} + \lambda_{cyc}\mathcal{L}_{cyc} \\
 &+ \lambda_{feat}(\mathcal{L}_{A~recon}^{feat} + \mathcal{L}_{B~recon}^{feat})\\
 &+ \lambda_{cont}(\mathcal{L}_{A~recon}^{c}+ \mathcal{L}_{B~recon}^{c}) \\
 &+  \lambda_{dom}(\mathcal{L}_{A~recon}^{d}+ \mathcal{L}_{B~recon}^{d}), \\
\end{aligned}
\end{equation}
where $\lambda_{cyc}$, $\lambda_{feat}$, $\lambda_{cont}$ and $\lambda_{dom}$ are hyperparameters weighted on each loss function. More details and codes can be found on the online link:  \href{https://github.com/vivivic/speech-domain-adaptation-DRL}{https://github.com/vivivic/speech-domain-adaptation-DRL}.

\subsection{Model configuration}
The proposed model is composed of a pair of discriminators, domain and context encoders, and decoders. Each discriminator or domain encoder is composed of a four-layer CNN followed by a four-layer FCNN. Each context encoder is a four-layer CNN. Each decoder has an initial four-layer FCNN to receive the domain latent variables as the affine parameters of adaptive instance normalization. Then, it has six residual blocks to perform the adaptive instance normalization. Finally, it has four upsampling layers to generate converted speech features. More details are summarized in Table~\ref{table1}.
\begin{table}
\caption{Model configuration.}
\label{table1}
\centering
\begin{tabular}{ll}
\toprule
\multicolumn{2}{c}{(kernel size) / (channel) / (stride)}\\
\midrule
     \textbf{discriminator} \\
    \midrule
    6x6 / 8 / 2x2  / conv  \\
    6x6 / 16 / 2x2 /  conv     \\
    1x6 / 32 / 1x2 / conv     \\
    1x3 / 64 / 1x2 /  conv    \\
    1600-512-256-64-1  dense layer        \\
    \midrule
     \textbf{domain encoder}&  \textbf{context encoder}\\
    \midrule
    1x6 / 8 / 1x2  / conv  & 1x6 / 16 / 1x2  / conv\\
    1x6 / 16 / 1x2 /  conv  &1x6 / 32 / 1x2  / conv   \\
    1x6 / 32 / 1x2 / conv    &1x6 / 64 / 1x2  / conv \\
    1x3 / 64 / 1x2 /  conv    &1x6 / 128 / 1x2  / conv\\
    256-128-32-16-8  dense layer   &     \\
\midrule
\textbf{decoder} \\
\midrule
\multicolumn{2}{l}{8-16-32-64-128 dense layer for affine parameters}\\
\multicolumn{2}{l}{6 x residual block(3x3, 128) with AdaIN }\\
    1x3 / 8 / 1x2  / Tconv  & \\
    1x3 / 16 / 1x2 /  Tconv  &\\
    1x6 / 16 / 1x2 / Tconv    & \\
    1x6 / 16 / 1x2 /  Tconv    &\\
\bottomrule
\end{tabular}
\end{table}

\section{Experiments}	
 \subsection{Noisy-to-clean speech domain adaptation}
We evaluated the performance of the proposed method using the CHiME4 challenge corpus~\cite{vincent2017analysis} and the Kaldi toolkit~\cite{povey2011kaldi}. The ASR system with deep-neural-network(DNN)-based acoustic model was built by basic Kaldi recipe for the CHiME4 challenge using 80-dimensional mel-spectral features. Spectral analysis was performed from 25-ms-long Hamming-windowed input speech at every 10ms. For our speech domain adaptation model, a 20-frame-long segment was used as an input. The hyperparameters $\lambda_{cyc}$, $\lambda_{feat}$, and $\lambda_{cont}$ were set to 1 whereas  $\lambda_{dom}$ was set to 5. The performance was measured by the word error rates (WERs).
The CHiME4 challenge corpus considered four different noisy environments (on a bus, cafe, pedestrian area, and street junction). In the corpus, the training set consisted of 1,600 real-recorded noisy speech utterances, and 7,138 simulated noisy speech utterances based on the clean utterances in the Wall Street Journal (WSJ0) SI-84 training set for multi-condition training. However, we used the original 7,138 clean utterances only to verify the performance of noisy-to-clean speech domain adaptation. We evaluated the performance of our method for 2,640 utterances in the evaluation set and also 3,280 utterances in the development set.

\begin{figure}
  \centering
  \includegraphics[width=1\columnwidth]{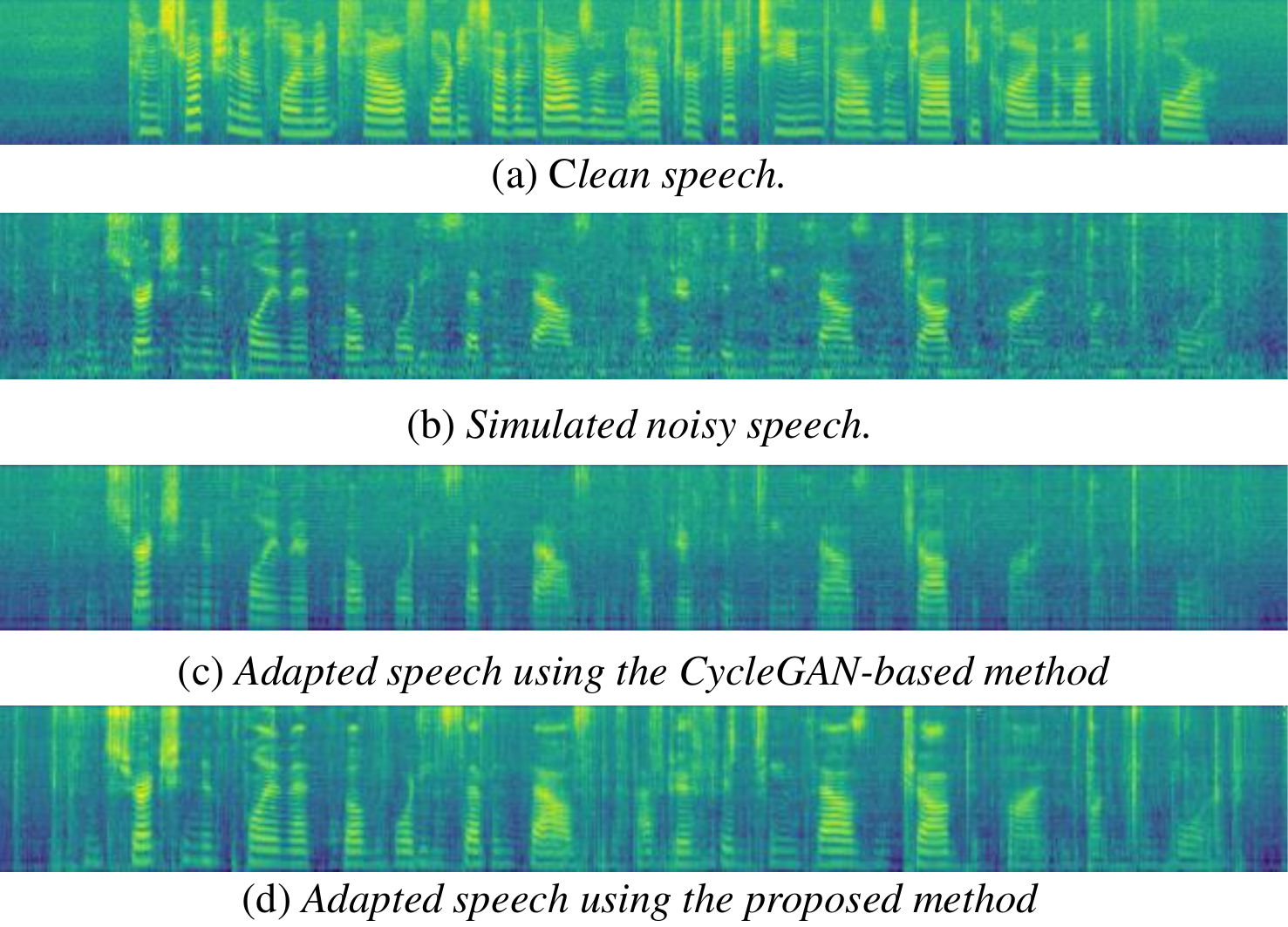}
  \caption{Mel-spectrograms of an example of noisy-to-clean speech domain adaptation for an utterance corrupted by the bus noise (050c0103.wav).}
  \label{figure3}
   \end{figure}
   
 Figure~\ref{figure3} displays mel-spectrograms for an example of noisy-to-clean speech domain adaptation for a simulated noisy utterance corrupted by the bus noise, and Table~\ref{table2} summarizes WERs. For comparison, we also implemented a speech domain adaptation method by directly applying CycleGAN with skip connections without the disentangled representation learning to speech as explained in Subsection 2.1. As shown in Fig.~\ref{figure3}, the CycleGAN-based method reduced noise components, but the harmonics and formant components of the speech were modified or eliminated. Therefore, the WERs of the CycleGAN-based method were higher than those of the baseline without any processing. On the other hand, our method effectively removed the background noise with the harmonic components and formants of the speech maintained, which resulted in revealing some harmonic components that were hard to see in the noisy speech. As shown in Table~\ref{table2}, the proposed method reduced the WERs averaged over four different noisy environments by 6.55\% for the simulated utterances and by 15.70\% for the real-recorded utterances in the development set, respectively. In the evaluation set, the averaged WERs were lower than those of the baseline by 6.72\% for the simulated utterances and by 10.88\% for the real-recorded utterances, respectively.

\begin{table}
\vspace{0.5cm}
\caption{WERs for noisy-to-clean speech domain adaptation with the CHiME4 challenge corpus.}
\label{table2}
\centering
\begin{tabular}{@{~}c@{~}c@{~~~}c@{~~~}c@{~~~}c@{~~~}c@{~~~}c}
\toprule
&&\multicolumn{5}{c}{\textbf{ WERs(\%)} }\\
\textbf{method}&\textbf{task type}&\textbf{avg}&\textbf{bus}&\textbf{caf}&\textbf{ped}&\textbf{str}\\
\midrule
\midrule
&\multicolumn{6}{c}{development set}\\
\midrule
\multirow{2}{*}{baseline}&simu&57.01 &50.49 &65.96 &49.19 &62.39\\
		   		        &real&58.77 &71.47 &63.30 &43.30 &57.03\\
\multirow{2}{*}{CycleGAN}&simu&66.99&65.94&69.47&64.38&68.17\\
		   			  &real&62.11&78.01&62.67&52.35&55.40\\
\multirow{2}{*}{proposed}&simu&\textbf{50.46}&\textbf{41.45}&\textbf{60.88}&\textbf{45.18}&\textbf{54.34}\\
		   			&real&\textbf{43.07}&\textbf{50.18}&\textbf{47.33}&\textbf{32.19}&\textbf{32.59}\\
\midrule
&\multicolumn{6}{c}{evaluation set }\\
\midrule
\multirow{2}{*}{baseline}&simu&72.50 &66.29 &76.69 &75.18 &71.85\\
		   			&real&81.09 &94.20 &83.49 &80.83 &65.84\\
\multirow{2}{*}{CycleGAN}&simu&86.21&90.21&85.26&85.88&83.88\\
		   			 &real&89.92&97.66&90.46&89.97&81.58\\
\multirow{2}{*}{proposed}&simu&\textbf{65.78}&\textbf{55.34}&\textbf{72.45}&\textbf{71.27}&\textbf{64.05}\\
		   			&real&\textbf{70.21}&\textbf{82.31}&\textbf{74.90}&\textbf{70.89}&\textbf{52.75}\\		    			
\bottomrule
\end{tabular}
\end{table}
 
\subsection{Speech gender adaptation}
In order to show the capability of speech-to-speech adaptation from a speaker to a target speaker, we also performed an experiment on gender-mismatched recognition. We used the TIMIT corpus composed of speech sentences uttered by 438 males and 192 females~\cite{garofolo1993timit}. The ASR systems with DNN-based acoustic models were trained for data uttered by either males or females. The used speech features were the same as before. To train the domain adaptation model, we used the same segments with the same hyperparameters as before except that  $\lambda_{dom}$ was set to 10.

\begin{figure}[t]
  \centering
  \includegraphics[width=1\columnwidth]{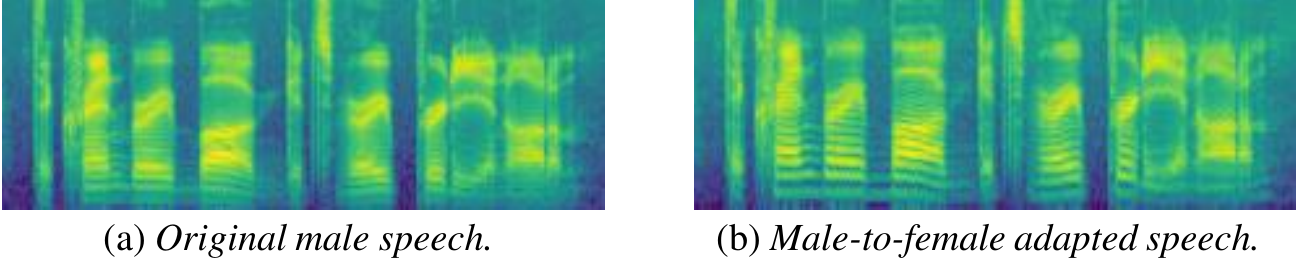}
  \caption{Mel-spectrograms of an example of male-to-female speech domain adaptation for an utterance (MBPM-SX317.wav).}
\label{figure4}
    \end{figure}

\begin{table}
\vspace{0.5cm}
 \caption{WERs for speech gender adaptation with the TIMIT corpus.}
  \label{table3}
  \centering
  \begin{tabular}{lllll}
     \toprule
     &&&\multicolumn{2}{c}{\textbf{ WERs(\%) } }\\
    \textbf{method}  & \textbf{training}&\textbf{test} & \textbf{dev} & \textbf{test} \\
    \midrule
    \midrule
    \multirow{2}{*}{}&female & female&22.5&20.8\\
		            &male & male&17.0&20.3\\
    \midrule
    \multirow{2}{*}{baseline}&female&male&33.3&35.9\\
		                          &male&female&31.8&29.8\\
    \multirow{2}{*}{CycleGAN}&female&male&\textbf{22.5}&\textbf{26.0}\\
		                          &male&female&\textbf{22.3}&\textbf{20.8}\\
    \multirow{2}{*}{Proposed}&female&male&23.9&29.0\\
		                          &male&female&23.6&21.7\\		
      \bottomrule
  \end{tabular}
\end{table}
 
Mel-spectrograms of an example of male-to-female speech conversion are shown in Fig.~\ref{figure4}, and the WERs were summarized in Table~\ref{table3}. In Fig.~\ref{figure4}, one may find the converted speech with the pitch frequency shifted upward. In Table~\ref{table3}, the proposed method improved the recognition performance with slightly higher WERs than the CycleGAN-based method because the CycleGAN-based method employed one simple encoder and was easier to train the model for conversion of speech itself than ours.

\section{Conclusions}
In this paper, we proposed the speech domain adaptation method based on GANs with disentangled representation learning, and evaluated it by the ASR experiments. To apply the disentangled representation learning, two separated encoders were used to generate latent variables with different attributes, and the adaptive instance normalization method was exploited to apply the latent variables to improve generation of adapted speech features. Our model did not require paired utterances and was simply applied on the mel-spectrograms of speech. The experimental results demonstrated that our model was effective in speech domain adaptation, especially for robust ASR. In the future, we will focus on noisy-to-clean speech conversion to achieve better robustness by modeling domain latent variables with Gaussian mixture models.

\section{Acknowledgment}
This work was supported and funded by Clova AI Research, Naver corporation.
\bibliographystyle{IEEEtran}
\bibliography{my}
\end{document}